\documentclass[aps,prl,showpacs,twocolumn,superscriptaddress,notitlepage]{revtex4-1}
\usepackage{amsmath, amssymb, amsthm}
\usepackage{mathtools}
\usepackage{mathrsfs}
\usepackage{bm}
\usepackage{slashed}   %feynman slashed symbols
\usepackage{graphicx}
\usepackage{multirow}
\usepackage{tikz}
\usepackage{subfigure}
\usepackage{relsize}	%rescalable math
\usepackage{array}
\usepackage{float}
\usepackage{color}
\usepackage{xcolor}
\usepackage{soul}
\usepackage{ulem}
\usepackage{lipsum}
 \usepackage{longtable}
\usepackage{hyperref}
\hypersetup{colorlinks=true,linkcolor=blue,anchorcolor=blue,citecolor=darkgreen, filecolor=blue,urlcolor=blue,bookmarksnumbered=true,
pdfview=FitB
}
\def\dd{{\mathrm{d}}}

\mathchardef\-="2D

\newcommand{\half}[1][1] {\mathsmaller{\frac{#1}{2}}}

\colorlet{darkgreen}{green!60!black}
\colorlet{brightyellow}{yellow!75!red}
\colorlet{orange}{red!50!yellow}
\colorlet{darkblue}{blue!60!black}
\colorlet{darkred}{red!80!black}
\colorlet{greenblue}{green!50!blue}

\makeatletter

\newcommand{\Rmnum}[1]{\expandafter\@slowromancap\romannumeral #1@}
\makeatother

\begin{document}
%opening
\title{Flavor asymmetry from the non-perturbative nucleon sea}

%first author
\author{Yihan Duan}
\affiliation{Department of Modern Physics, University of Sciences and Technology of China, Hefei 230026, China}

\author{Siqi Xu}
\affiliation{Institute of Modern Physics, Chinese Academy of Sciences, Lanzhou 730000, China}
\affiliation{School of Nuclear Science and Technology, University of Chinese Academy of Sciences, Beijing 100049, China}
\affiliation{Department of Physics and Astronomy, Iowa State University, Ames, IA 50011, USA}

\author{Shan Cheng}
\affiliation{School of Physics and Electronics, Hunan University, 410082 Changsha, China}

\author{Xingbo Zhao}
\affiliation{Institute of Modern Physics, Chinese Academy of Sciences, Lanzhou 730000, China}
\affiliation{School of Nuclear Science and Technology, University of Chinese Academy of Sciences, Beijing 100049, China}
\affiliation{CAS Key Laboratory of High Precision Nuclear Spectroscopy, Institute of Modern Physics, Chinese Academy of Sciences, Lanzhou 730000, China}

%correponding author
\author{Yang Li}
%\email{leeyoung1987@ustc.edu.cn} 
\thanks{Corresponding author}
\affiliation{Department of Modern Physics, University of Sciences and Technology of China, Hefei 230026, China}
\affiliation{Anhui Center for Fundamental Sciences in Theoretical Physics, University of Science and Technology of China, Hefei, 230026, China}

\author{James P. Vary}
\affiliation{Department of Physics and Astronomy, Iowa State University, Ames, IA 50011, USA}

\date{\today}
\begin{abstract}

We demonstrate, in the context of a scalar version of the chiral effective field theory, that the multi-sea quark contribution to the nucleon is significant and highly non-trivial in sharp contrast to the prediction of perturbation theory. The non-perturbative calculation is performed in the Fock sector dependent renormalization scheme on the light front, in which the non-perturbative renormalization is incorporated. The calculation suggests that a fully non-perturbative calculation of the chiral EFT is needed to obtain a robust result to be compared with the recent experimental measurement of flavor asymmetry in the proton. 

\end{abstract}

\maketitle

\paragraph{Introduction}
Understanding the sea-quark contribution of hadrons is of fundamental importance \cite{Kumano:1997cy, Garvey:2001yq, Chang:2014jba, Geesaman:2018ixo}. It not only provides a sharper view of the fundamental structure of matter \cite{Gottfried:1967kk}, but also serves as a tool for detecting physics beyond Standard Model \cite{NNPDF:2021njg}. Recently, the SeaQuest/E906 experiment published their first measurement of the flavor asymmetry $\bar d/\bar u$ of the proton sea from the muon pair production within the Drell-Yan process of high-energy proton-ion collisions \cite{SeaQuest:2021zxb, Kotlorz:2021goc, FNALE906:2022xdu}, challenging the previous measurement by the NuSea/E866 experiment at large momentum fraction $x \gtrsim 0.3$ \cite{NA51:1994xrz, HERMES:1998uvc, NuSea:1998tww, NuSea:1998kqi, NuSea:2001idv, Salajegheh:2017iqp}. 
In particular, the NuSea/E866 experiment predicted that $\bar d/\bar u$ dips below unity at $x \sim 0.3$, which is contrary to the SeaQuest/E906 measurement. 
The tension between these experimental measurements calls for a better understanding from theories including global analysis \cite{Harland-Lang:2014zoa, Alekhin:2017kpj, Accardi:2019ofk, NNPDF:2021njg, Cocuzza:2021cbi, Hou:2022onq, Cocuzza:2022jye}. It was generally believed that the flavor asymmetry of the proton encodes the non-perturbative physics of QCD, making it a particular challenge for calculations from first principles \cite{Scapellato:2019csz, Ji:2020ect}. 
The flavor asymmetry is also related to other sea quark asymmetries, such as the nucleon strangeness as well as the intrinsic charm \cite{Brodsky:1996hc}.

Among various theoretical interpretations \cite{Henley:1990kw, Dorokhov:1993fc, Steffens:1996bc, Londergan:1998ai, Miller:2002ig, Trevisan:2008zz, Song:2011fc, Kamano:2012yx, Burkardt:2012hk, Ji:2013bca, Wang:2014sua, Ehlers:2014jpa, Salamu:2014pka, Lin:2014zya, Basso:2015lua, Kofler:2017uzq, Tadepalli:2019ixa, Bourrely:2019ysc, Soffer:2019gbb,  Leon:2020bvt, Alexandrou:2021oih, Wang:2021ild, Alberg:2021nmu, Signal:2021aum, Cui:2021gzg, Kotlorz:2021goc, Chang:2022jri, Leinweber:2022guz, He:2022leb, Bourrely:2022mjf, Wang:2022bxo, Bonvini:2023tnk, Choudhary:2023unw, Choudhary:2023bap}, the pion cloud model \cite{Theberge:1980ye, Thomas:1981vc, Thomas:1983fh} has shown a trend consistent with the SeaQuest/E906 results as well as other theories at small and modest $x$. It can also be used to investigate the sea quark-antiquark asymmetry \cite{Brodsky:1996hc, Sufian:2018cpj, Sufian:2020coz}.
In this model, the sea quark contribution is viewed as the hadron cloud around the nucleon. The interaction  between the pion cloud and the host nucleon is described by chiral effective field theory ($\chi$EFT). Recently, Alberg~et.~al. showed that light-cone methods can be used to compute the pionic longitudinal momentum distribution function (LMDF) $f_{\pi N}(x)$ \cite{Alberg:2012wr, Alberg:2017ijg, Alberg:2021nmu}. Then, the quark distribution function can be represented as, 
%\begin{widetext}
\begin{multline}\label{eqn:convolution} 
q_N(x) = Z q_{N}^{(0)}(x) + 
  \sum_{B}  \int_x^1 \frac{\dd y}{y} \big[ f_{\pi B}(y) q_{\pi}(x/y)  \\
 + f_{B\pi}(y) q_{B}(x/y)\big]\,.
\end{multline}
%\end{widetext}
Alberg et. al.'s model is successful in describing the nucleon sea at small and modest $x$. At large $x\gtrsim 0.3$, this model predicts $\bar d/\bar u$ is always above unity. However, the results start to show deviation from the experimental data of SeaQuest/E906 \cite{Alberg:2017ijg}. The predictions are improved after taking into the evolution of the sea \cite{Alberg:2021nmu}. 
Alberg~et.~al.'s calculation is based on leading-order (LO) light cone perturbation theory which is equivalent to Fock sector expansion up to the one pion sector: $|N\rangle_\mathrm{phy} = |N\rangle + |N\pi \rangle + |\Delta \pi\rangle$. A non-perturbative extension in a limited Fock space involving a single pion has also appeared~\cite{Du:2019ips}. One of the immediate questions is how large are the non-perturbative effects such as multi-pion contributions. Indeed, the effective coupling, $g_{\pi N}\simeq 13$, is large.  
Furthermore, the pion carrying large momentum fraction $y$ is ultra-relativistic and its splitting function is more likely modified in the non-perturbative regime, which directly affects the parton distribution functions (PDFs) at large $x$.

In this work, we investigate the non-perturbative effects in the pion cloud model by exploiting the recent success in solving strongly coupled scalar field theories in the light-front Hamiltonian formalism \cite{Li:2015iaw, Karmanov:2016yzu}.  
We perform a systematic Fock expansion and incorporate the multi-pion component of the nucleon sea, 
$|N\rangle_\text{ph} = |N\rangle + |N\pi\rangle +  |\Delta \pi\rangle + |N\pi\pi\rangle +  |\Delta \pi\pi\rangle$, where the Fock sector expansion is observed to be converged \cite{Li:2015iaw}. 
We show that the multi-pion sea modifies the prediction for the flavor asymmetry in the region from moderate to large coupling that is relevant for the pion cloud model. 
The adoption of the scalar theory certainly simplifies the non-perturbative calculation at the cost of major differences from $\chi$EFT. These differences are reflected in the pion longitudinal momentum distribution at small $x$, which are shown in this work. We instead focus on the non-perturbative effects within the present model. Our results suggest that a full non-perturbative calculation starting from the $\chi$EFT is needed to obtain a robust result to be compared with the recent experimental measurement of the flavor asymmetry in the nucleon. 

\paragraph{Formalism}

We follow Alberg~et.~al.'s treatment of the pion cloud model \cite{Alberg:2021nmu}, except starting from scalar type pion-nucleon and pion-nucleon-$\Delta$ interactions, viz.,
\begin{equation}
\mathscr L_\text{int} = g_{N\pi} \overline N N \pi + g_\Delta \overline N \Delta \pi + g_\Delta \overline \Delta N \pi,
\end{equation}
where, $N, \Delta$ are complex scalar fields and $\pi$ is a real scalar field. Note that the couplings $g_{N\pi}$ and $g_\Delta$ have the dimension of mass. A dimensionless coupling can be introduced, $\alpha = g^2_{N\pi}/(16\pi M^2_N)$.   
The theory is a low-energy approximation of $\chi$EFT applicable to moderate and large $x$.  It still encapsulates some general features of the nucleon sea. Comparing the LO perturbative results for LMDF from $\chi$EFT \cite{Alberg:2021nmu} and from the scalar Yukawa theory, we find that they are almost identical except for some function that can be absorbed into phenomenological form factors. In this work, we focus on the non-perturbative effects within the same model and adopt this simplified model. 

\begin{figure}
\centering
\includegraphics[width=0.45\textwidth]{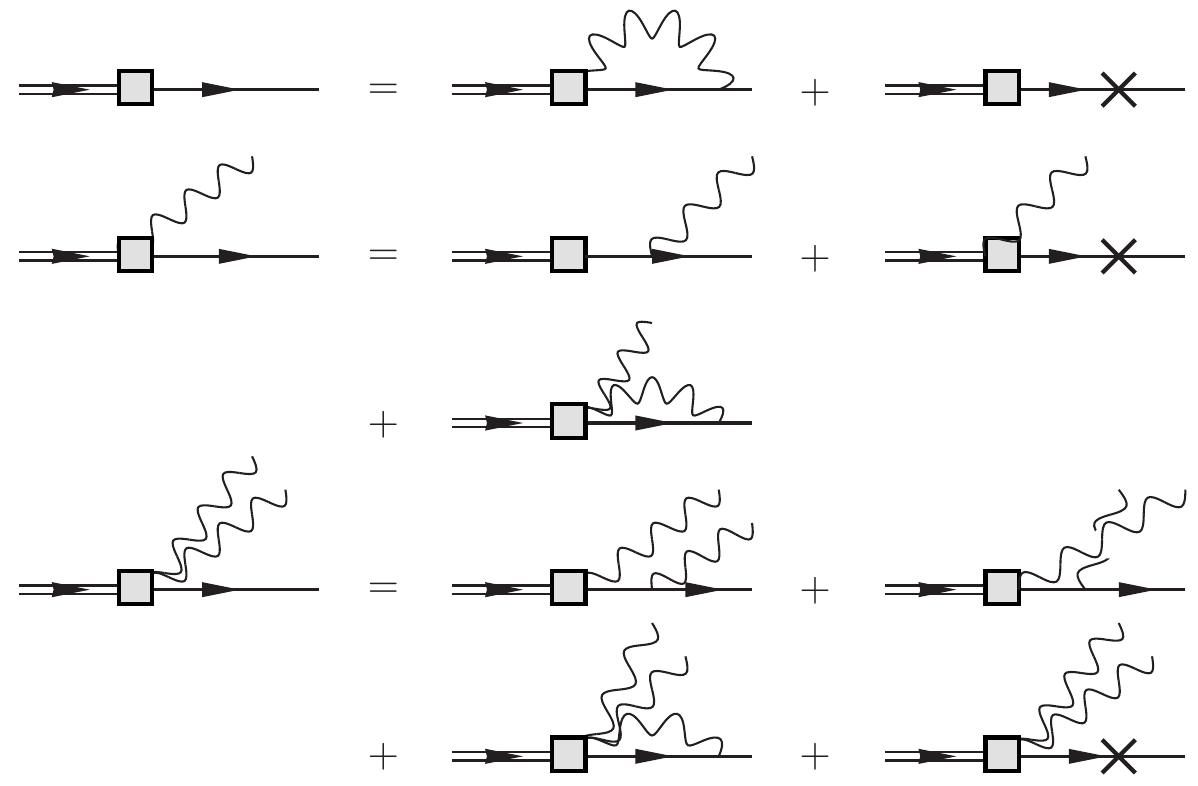}
\includegraphics[width=0.45\textwidth]{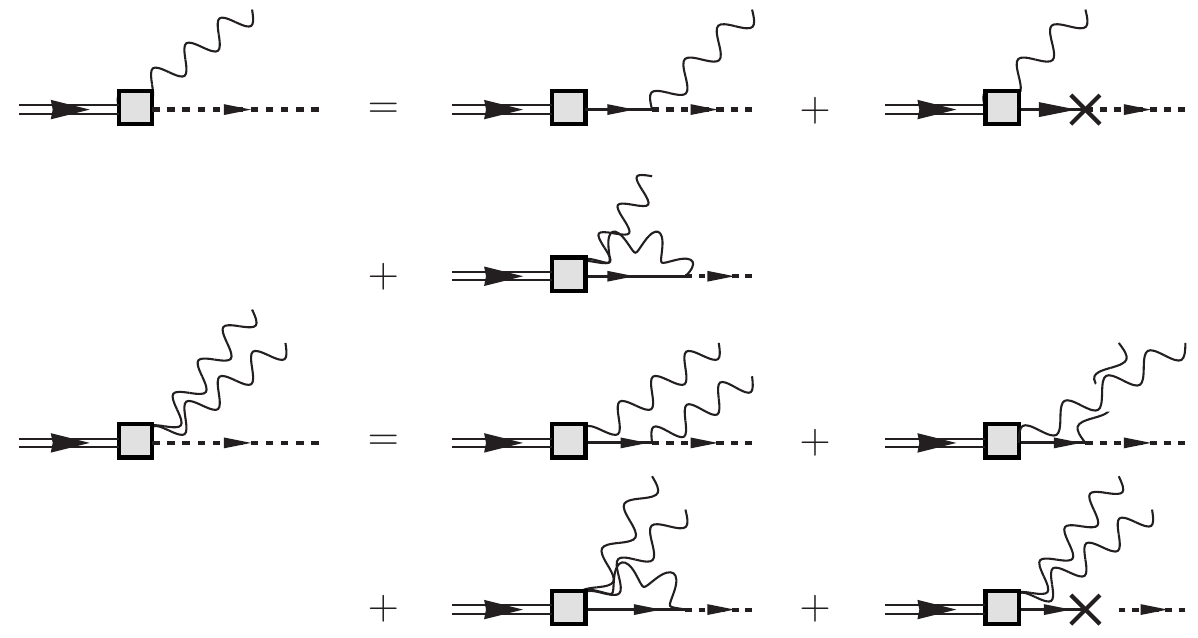}
\caption{Diagrammatic representation of the system of equations with a truncation up to two pions which is a subset of the three pion equations that we employ here. The solid lines represent the (mock) bare nucleons. The dashed lines represent the (mock) $\Delta$'s. 
And the wavy lines represent the (mock) pions. The double solid lines represent the physical nucleons. }
\label{fig:eigenvalue_equation_diagrams}
\end{figure}

To obtain the state vector $|N\rangle_\text{phy}$ of the physical nucleon, we diagonalize the light-front quantized Hamiltonian in Fock space. 
See Refs.~\cite{Pauli:1985pv, Pauli:1985ps, Hornbostel:1988fb, Tang:1991rc, Brodsky:1991ir, Brodsky:1997de, Vary:2009qz, Vary:2009gt} for reviews of non-perturbative methods developed to diagonalize the light-front Hamiltonian. Fig.~\ref{fig:eigenvalue_equation_diagrams} shows the first few diagrams of the system of equations derived from the light-front Schrödinger equation, which are an infinite set of coupled integral equations in Fock space. A review of the relevant Feynman rules is collected in Refs.~\cite{Brodsky:1997de, Carbonell:1998rj}. Note that the non-perturbative vertex functions (shaded blocks) are employed to extend the Feynman rules to the non-perturbative regime. The vertex functions are the $T$-matrix elements and are related to the light-front wave functions by resolvents \cite{Hiller:1998cv}. 
The explicit expressions translated from these diagrams are shown in Refs.~\cite{Li:2015iaw, Karmanov:2016yzu}. 

To keep the system of equations numerically tractable, we impose the Fock sector truncation, and investigate the convergence of observables with respect to the Fock sector expansion \cite{Hiller:2016itl}. The truncation up to two-body ($|N\rangle+|N\pi\rangle$) is identical to the LO perturbation theory. 
In Ref.~\cite{Li:2015iaw}, the state vector of the physical nucleon was solved up to four-body truncation ($|N\rangle+|N\pi\rangle+|N\pi\pi\rangle+|N\pi\pi\pi\rangle$) and it was shown that the Fock sector expansion is well converged up to three-body ($|N\rangle+|N\pi\rangle+|N\pi\pi\rangle$). We thus adopt the four-body truncation result as the non-perturbative results and estimate the uncertainty associated with the Fock sector truncation as the difference between the three-body truncation and the four-body truncation. 
The Fock states that contains $\Delta$, i.e. $|\Delta\pi\rangle+|\Delta\pi\pi\rangle+|\Delta\pi\pi\pi\rangle$, are obtained from the nucleon wave functions as shown in Fig.~\ref{fig:eigenvalue_equation_diagrams}. Effectively, a small back reaction of $\Delta$ on the nucleon and pion Fock states is neglected for numerical simplicity. Note that we excluded sub-diagrams that convert a nucleon to a $\Delta$ or vice versa since such a process violates the angular momentum conservation. 

While the scalar theory in (3+1)D is super-renormalizable, it still contains divergences. The divergences can be absorbed into the bare parameters by imposing the mass and coupling constant renormalization conditions \cite{Li:2015iaw}. For the mass renormalization, we require the mass eigenvalue equals the physical mass. For the coupling constant renormalization, we require the $T$-matrix $\Gamma_{N\pi/N}$ equals the physical coupling multiplied by the necessary $Z$ factors at the on-shell point. 
 The whole procedure is non-perturbative, and is the standard one, to be distinguished with the perturbative treatment in Ref.~\cite{Alberg:2021nmu}.  N.B. the values of the bare parameters, i.e., the bare mass $m_0^2$ and the bare couplings $g_0$, depend on the Fock sector truncation. Fortunately, boost invariance in light-front dynamics guarantees that these sector dependent counter-terms can be obtained sector by sector recursively. This renormalization scheme is known as the Fock sector dependent renormalization \cite{Karmanov:2008br}, which has been successful in renormalizing the light-front Hamiltonian operators of the scalar theory \cite{Li:2015iaw, Karmanov:2016yzu, Cao:2023ohj}, Yukawa theory \cite{Karmanov:2010ih, Karmanov:2012aj} and QED \cite{Chabysheva:2009vm, Chabysheva:2009ez} as well as $\chi$EFT type interactions \cite{Tsirova:2010zza}. 

\begin{figure}
\centering
\includegraphics[width=0.25\textwidth]{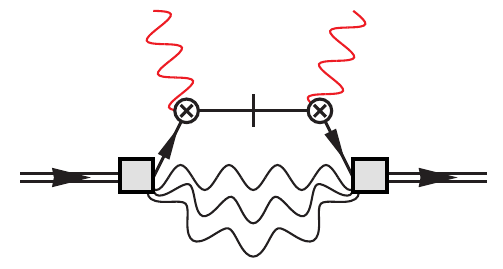}
\caption{The light-front wave function representation of the longitudinal momentum distribution's contribution to deep inelastic scattering as depicted by the handbag diagram corresponding to the contributions from the four-body sector to the physical nucleon.}
\label{fig:handbag}
\end{figure}

\paragraph{Longitudinal momentum distribution functions (LMDF)}

The PDF describes the probability of finding a collinear parton with fractional momentum $x$ inside the proton. 
In analogy to the QCD definition, we introduce a LMDF as \cite{Diehl:2003ny},
\begin{multline}
 f_{B\pi} (x) = \frac{k^+}{2} \int \dd z^- e^{\half[i]xp^+z^-} \\
 \times \langle p|\bar B(-\half[z]) B(\half[z])|p\rangle\big|_{z^+ = z_\perp = 0}.
\end{multline}
where $x = {k^+}/{p^+}$ is the longitudinal momentum fraction of the baryon $B$ inside the physical nucleon, $B = N, \Delta$. This expression is diagrammatically shown in Fig.~\ref{fig:handbag}.
In the light-front wave function representation \cite{Diehl:2000xz}, 
\begin{equation}
f(x) = f_1(x) + f_2(x) + \cdots 
\end{equation}
where, the $n$-body contribution is (the $n$-th particle is taken as the baryon)
%\begin{widetext}
\begin{multline}
f_n(x) = \frac{1}{(n-1)!} \prod_{i=1}^n\int \frac{\dd x_i\dd^2k_{i\perp}}{(2\pi)^32x_i} 2(2\pi)^3 \delta (x - x_n) \\
\times \delta(x_1+\cdots+x_n-1) 
 \delta^2(\vec k_{1\perp}+\cdots+\vec k_{n\perp}) \\
\times  \big|\psi_n(x_1,\vec k_{1\perp}, x_2, \vec k_{2\perp}, \cdots, x_n, \vec k_{n\perp})\big|^2\,.
\end{multline}
%\end{widetext}
Since the scalar theory is super-renormalizable, the ``factorization scale'' of the LMDFs can be taken to infinity \cite{Aslan:2022zkz}.
The PDFs of the physical nucleon are obtained from convoluting the LMDF with the PDFs of its constituent hadrons as shown in Eq.~(\ref{eqn:convolution}). 
We adopt the same pion and bare proton PDFs as Alberg et. al. at a scale $\mu^2 = 25.5\,\mathrm{GeV}^2$ \cite{Alberg:2021nmu, Aicher:2010cb, Gluck:1999xe, Szczurek:1997fr}. 

\begin{figure}
\centering
\includegraphics[width=0.4\textwidth]{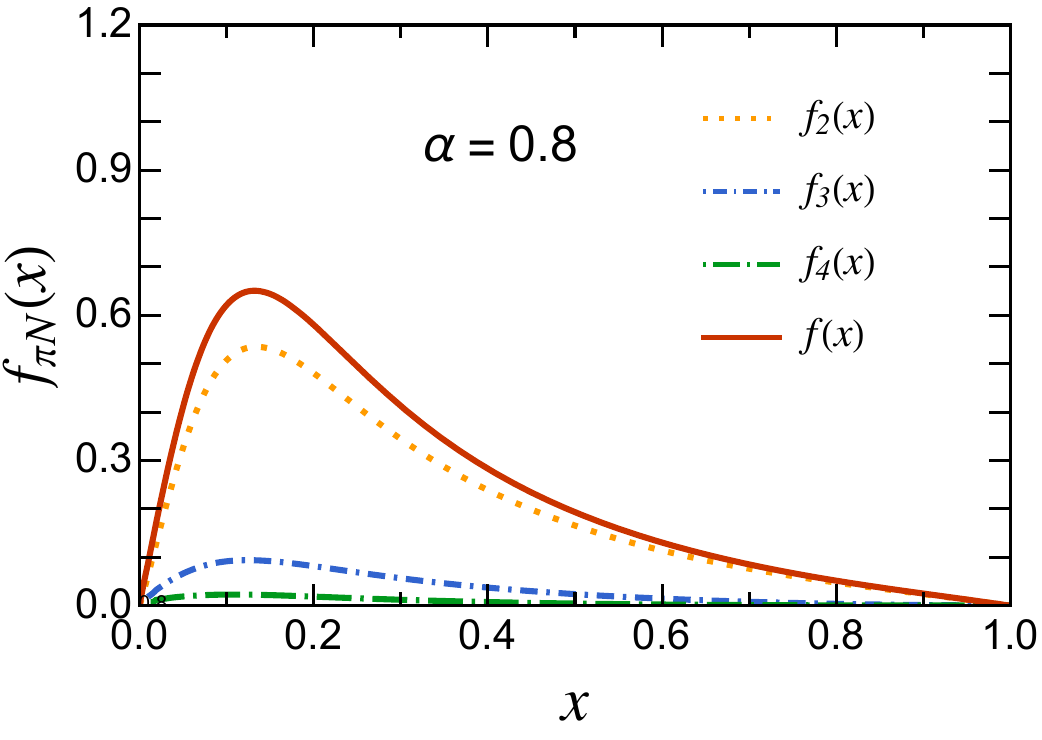}
\includegraphics[width=0.4\textwidth]{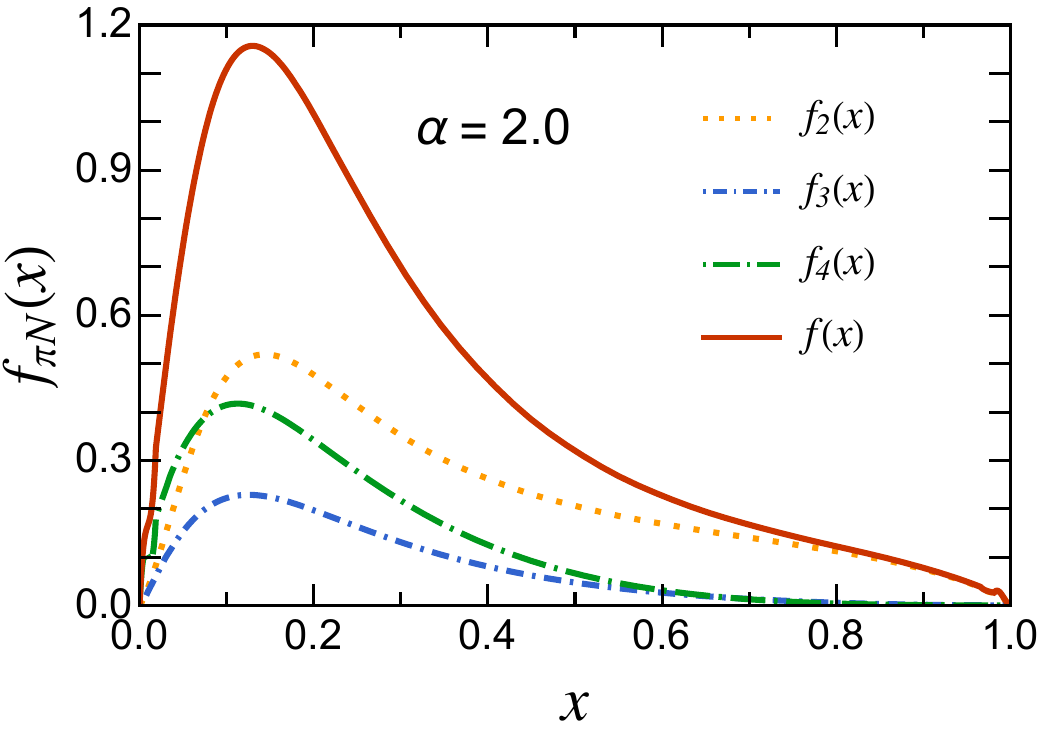}
\caption{(Colors online) Longitudinal momentum distribution function of the pion inside the physical scalar nucleon for coupling (\textit{Top}) $\alpha = 0.8$ and (\textit{Bottom}) $\alpha = 2.0$. The coupling to $\Delta$ is turned off in both cases, i.e. $g_\Delta = 0$. The contributions from the multi-pion Fock sectors are also shown in each panel. For $\alpha=2.0$ there are significant three- and four-body contributions at the large longitudinal momentum fraction $x$.}
\label{fig:LMDF}
\end{figure}

To demonstrate the effects of the multi-pion sea, we show in Fig.~\ref{fig:LMDF} the LMDFs of the pion inside the physical nucleon for two non-perturbative couplings, $\alpha = 0.8$ and $\alpha = 2.0$, up to one nucleon and three pions. We also turned off the coupling to $\Delta$, i.e. setting $g_\Delta = 0$, to focus on the multi-pion sea. 
Contributions from different Fock sectors are also shown. Comparing these LMDFs, as may be anticipated, the contribution from the multi-pion Fock sector increases with the increase in the coupling.  

\begin{figure}
\centering
\includegraphics[width=0.45\textwidth]{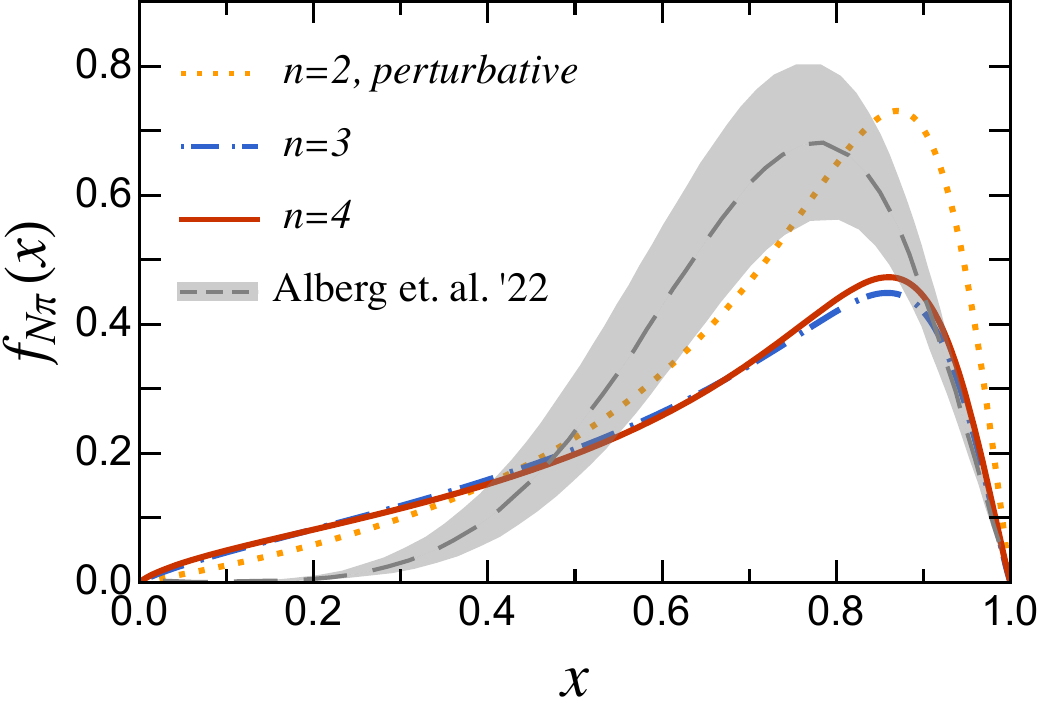}
\includegraphics[width=0.45\textwidth]{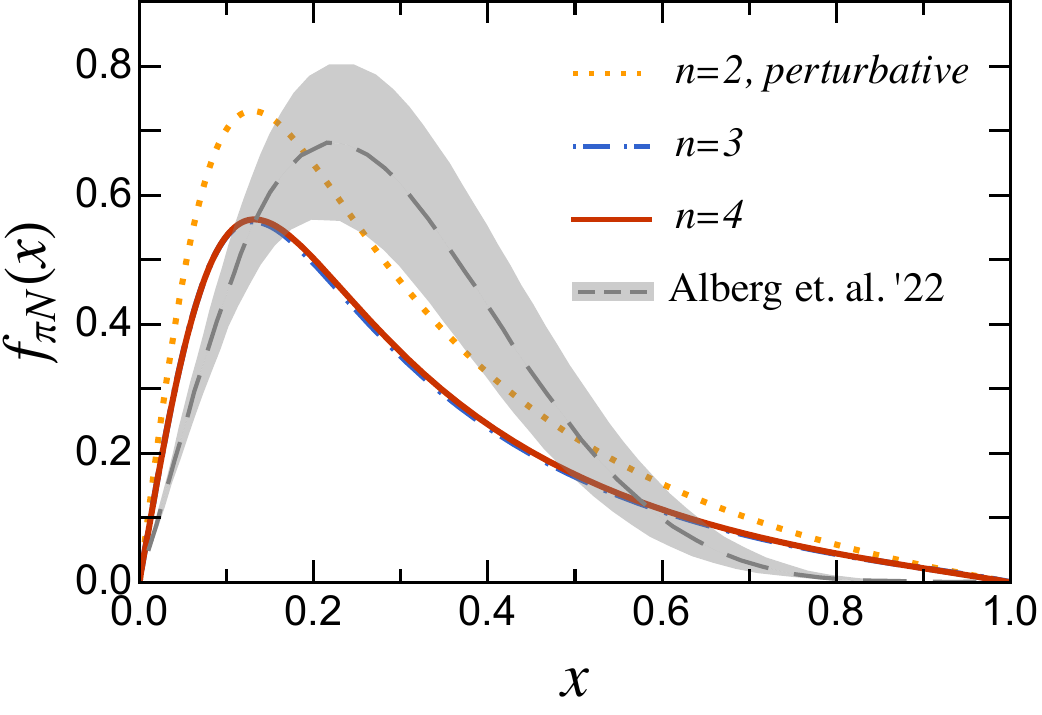}
\includegraphics[width=0.45\textwidth]{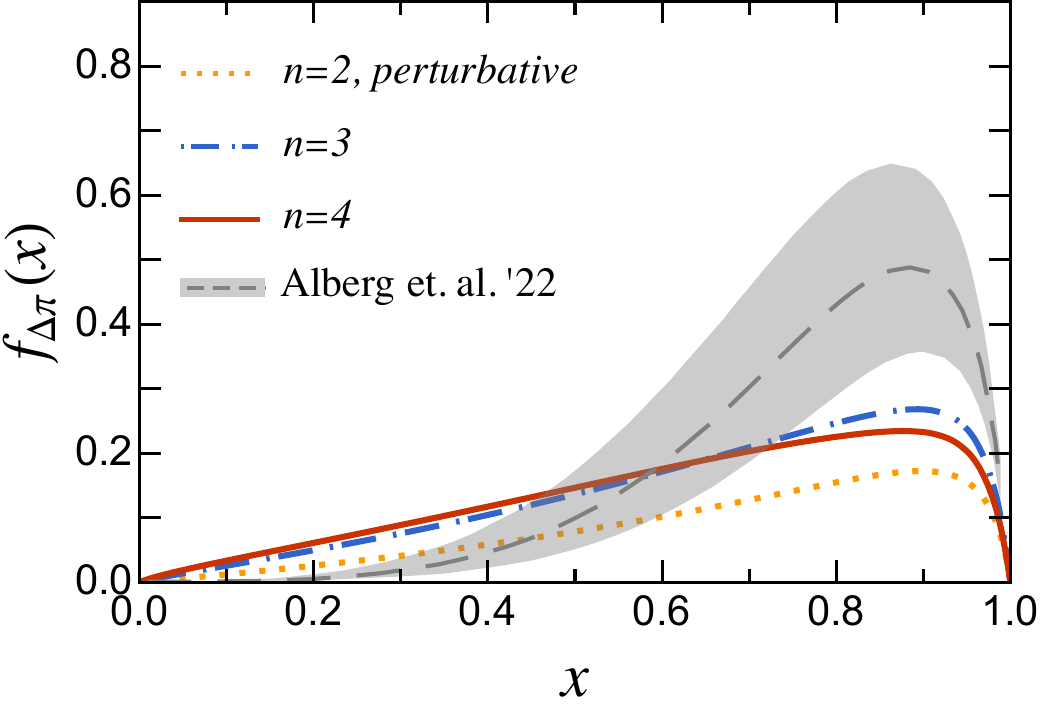}
\caption{(Colors online) Longitudinal momentum distribution function of the nucleon (\textit{Top}),  the pion (\textit{Middle}) and the $\Delta$ (\textit{Bottom}) inside the physical nucleon. Results from Fock sector truncation up to one baryon plus one pion ($n=2$), one baryon plus two pions ($n=3$) and one baryon plus three pions ($n=4$) are compared. The two-body truncation is equivalent to LO light-cone perturbation theory. Alberg et. al.'s results using LO perturbation theory with a chiral Lagrangian are included for comparison \cite{Alberg:2021nmu}.}
\label{fig:LMDF_N_Delta}
\end{figure}

\begin{figure}
\centering
\includegraphics[width=0.45\textwidth]{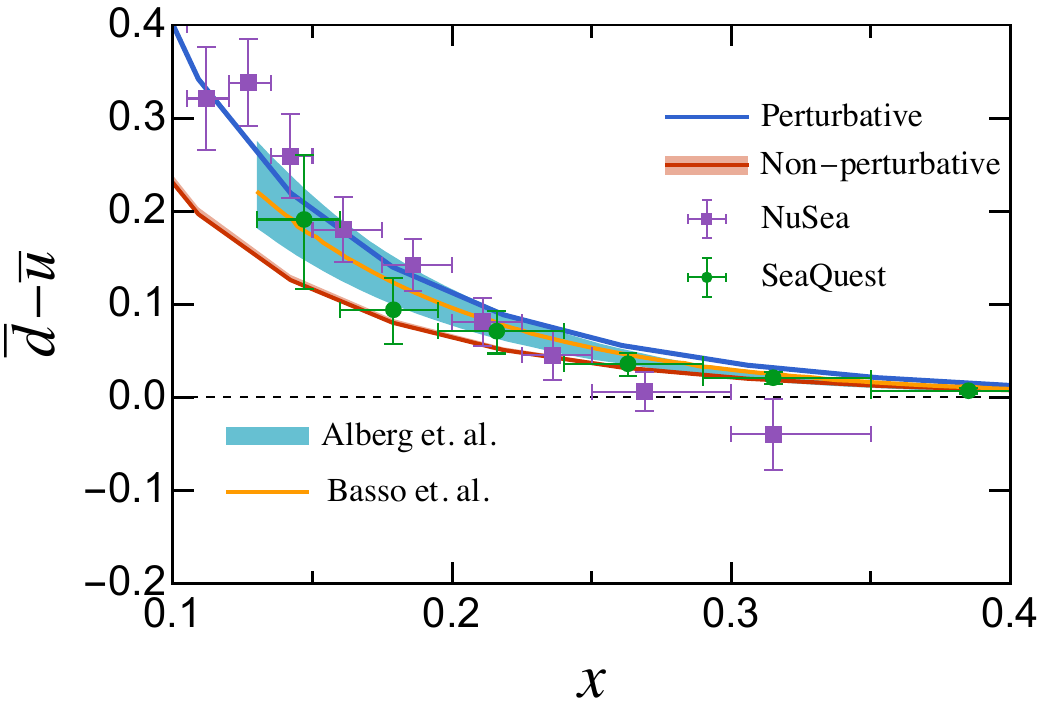} \\
\includegraphics[width=0.45\textwidth]{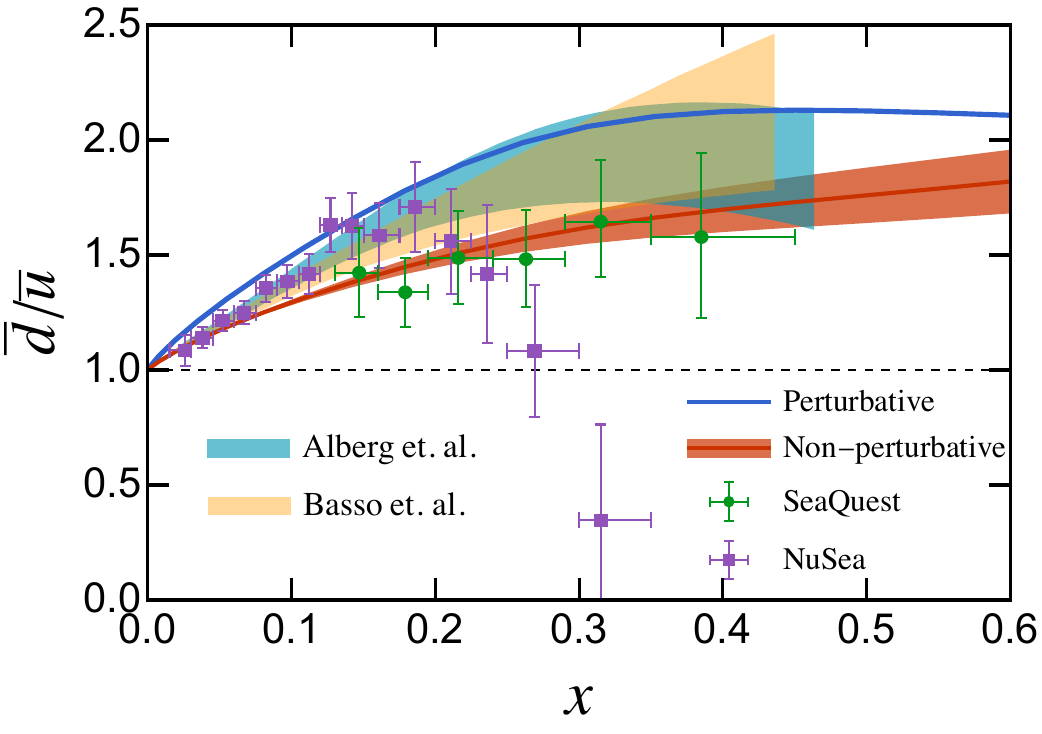}
\caption{Flavor asymmetry of the nucleon sea: (\textit{Top}) $\bar d - \bar u$; (\textit{Bottom}) $\bar d/\bar u$ from the perturbative and non-perturbative solutions of a pion cloud model with scalar type interactions in comparison with available experimental data from NuSea/E866 \cite{NuSea:1998kqi, NuSea:2001idv, Salajegheh:2017iqp} and SeaQuest/E906 \cite{SeaQuest:2021zxb, FNALE906:2022xdu} at the scale $Q^2 = 25.5 \,\mathrm{GeV}^2$.  
The non-perturbative result is obtained with a Fock space truncation up to four-body (one baryon plus up to three pions). 
The difference between the results with three-body and four-body truncations are shown as bands to indicate the convergence of the Fock space expansion. Results from the pion cloud model with light-cone perturbation theory by Alberg et. al. \cite{Alberg:2021nmu} and from the statistical model by Basso et. al. \cite{Basso:2015lua} are also shown for comparison. 
}
\label{fig:flavor_asymmetry}
\end{figure}

 It is tempting to use nonrelativistic reduction to relate $g_{\pi N} \approx 13$ to the couplings in the scalar theory. However, this procedure produces couplings well beyond the critical coupling for Landau poles \cite{Li:2015iaw}.  To compare with the Drell-Yan data, we adopt $g_{N\pi} = 6.0\,\mathrm{GeV}$ ($\alpha = 0.8$) and $g_\Delta =6.5\,\mathrm{GeV}$. These parameters are chosen to reproduce the SeaQuest $\bar d - \bar u$ data. Fig.~\ref{fig:LMDF_N_Delta} shows the LMDFs of the nucleon, the pion and the $\Delta$ obtained from the two-body, three-body and four-body truncations. As we mentioned, the two-body truncation is equivalent to the light-cone perturbation theory. The results obtained from three- and four-body truncations are close to each other, indicating a trend of numerical convergence of the Fock sector expansion.  Thereafter, we will adopt the four-body truncation as the non-perturbative result, and estimate the uncertainty using the difference between the three-body and four-body truncations. Alberg~et.~al.'s perturbative results based on a chiral Lagrangian are also shown in Fig.~\ref{fig:LMDF_N_Delta} for comparison \cite{Alberg:2021nmu}.  In general, our perturbative results differ from their perturbative results since we adopt the scalar version of the theory. Nevertheless, both the perturbative results of Ref.~\cite{Alberg:2021nmu} and our non-perturbative results reproduce the SeaQuest data $\bar d - \bar u$ as shown in Fig.~\ref{fig:flavor_asymmetry}. The integrated flavor asymmetry $\int_{0.13}^{0.45} \dd x [\bar d(x) - \bar u(x)] = 0.0122(7)$ is consistent with the SeaQuest/E906 measurement 0.0159(60) within the range $0.13 < x < 0.45$ \cite{FNALE906:2022xdu}.
 %. 

Figure~\ref{fig:flavor_asymmetry} shows the flavor asymmetry of the ``scalar nucleon" sea from this theory along with available experimental data at the scale $Q^2 = 25.5 \,\mathrm{GeV}^2$. Results from the pion cloud model by Alberg et. al. \cite{Alberg:2021nmu} and from the statistical model by Basso et. al. \cite{Basso:2015lua} are also shown for comparison. 
While perturbative and non-perturbative results for $\bar d - \bar u$ are close to each other, their difference in $\bar d/\bar u$ is moderately large, which signifies the high Fock sector contributions in the non-perturbative regime especially at large $x$.
 Within the same theory, our three- (one baryon plus up to two pions) and four-body (one baryon plus up to three pions) truncation results show a good convergence pattern as seen in Fig.~\ref{fig:LMDF_N_Delta}. However, these results are different from the two-body truncation results. The flavor asymmetry results exhibit significant sensitivity to the 2-body versus 4-body truncation as shown in Fig.~\ref{fig:flavor_asymmetry}, which signals important non-perturbative effects originating from the multi-pion sea contributions.

\begin{figure}
\centering
\includegraphics[width=0.45\textwidth]{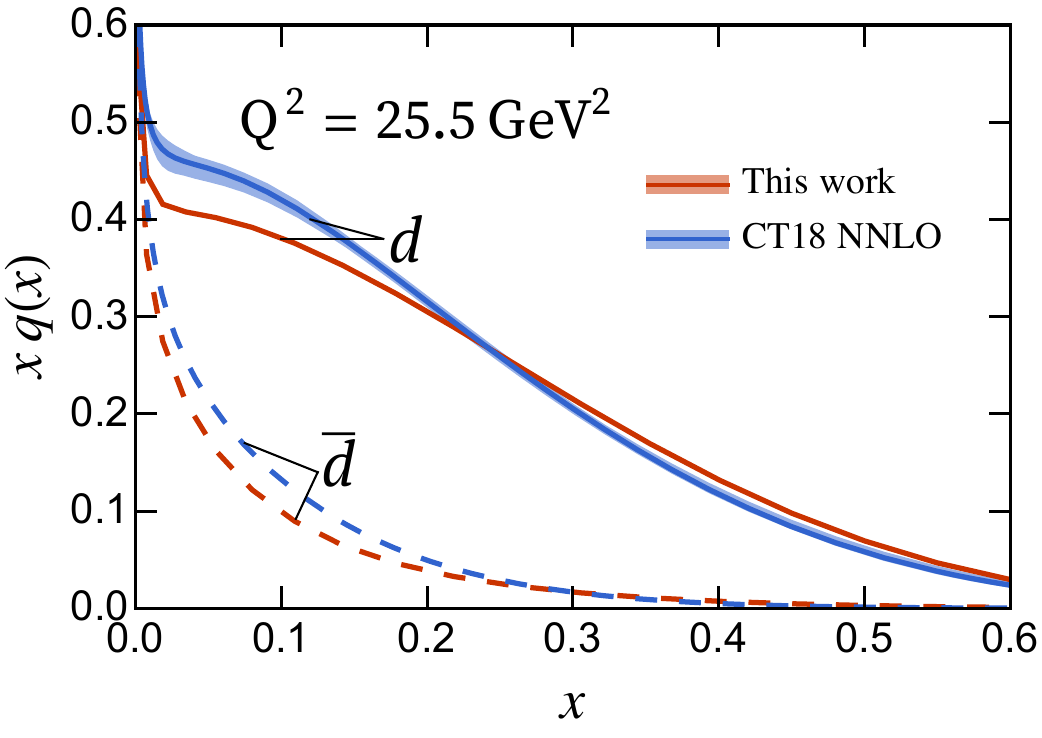}
\caption{The quark-antiquark asymmetry $d(x)$ vs $\bar d(x)$ obtained from this model using the non-perturbative solutions ($n=3,4$) of the scalar Yukawa model. The $n=3$ and $n=4$ results overlap with each other. The CT18 NNLO results are also shown for comparison \cite{Hou:2019efy}. 
}
\label{fig:ddbar_asymmetry}
\end{figure}

Finally, as a consistency check, Fig.~\ref{fig:ddbar_asymmetry} shows the quark-antiquark asymmetry $d(x)$ vs $\bar d(x)$ obtained from this model using the non-perturbative solutions. The CT18 NNLO results are shown for comparison \cite{Hou:2019efy}. 

\paragraph{Summary and outlook}

We investigated the non-perturbative effects of the nucleon sea within a scalar version of the pion cloud model. The non-perturbative dynamics is generated by the light-front Schrödinger equation with a systematic Fock sector expansion. Within this simple model, the convergence is observed already with three-body contributions (one baryon plus two pions) though we proceed to include four-body contributions as well. We show that the obtained longitudinal momentum distributions from the non-perturbative solutions are considerably different from  those obtained from a leading order light-front perturbation theory. The difference can be attributed to the multi-pion dynamics, which are the sea contributions within the pion cloud model. We also applied the solutions to describe the nucleon flavor asymmetry using the pion cloud model. While both the perturbative and non-perturbative sea can describe the $\bar d - \bar u$ data well, the non-perturbative nucleon sea can describe the $\bar d/\bar u$ data better for $x\gtrsim 0.1$. These comparisons between the perturbative and non-perturbative pion clouds indicates that the non-perturbative effects are important and may help improve the model. Indeed, part of the multi-pion sea can be taken into account by evolving the sea, which may help to explain the improved agreement in their updated prediction \cite{Alberg:2021nmu}. Nevertheless, a full investigation of the non-perturbative sea requires a non-perturbative calculation of  chiral effective field theory \cite{Tsirova:2010zza}, which is a natural next step.

\paragraph*{Acknowledgements}

The authors acknowledge valuable discussions with W. Du, V.A. Karmanov and P. Maris. We thank the referee for the discuss on the low-$x$ region.
This work was supported in part by the National Natural Science Foundation of China (NSFC) under Grant No.~12375081, by the Chinese Academy of Sciences under Grant No.~YSBR-101, and by the US Department of Energy (DOE) under Grant No. DE-SC0023692. 
X. Z. is supported by new faculty startup funding by the Institute of Modern Physics, Chinese Academy of Sciences, by Key Research Program of Frontier Sciences, Chinese Academy of Sciences, Grant No.~ZDBS-LY-7020, by the Natural Science Foundation of Gansu Province, China, Grant No.~20JR10RA067, by the Foundation for Key Talents of Gansu Province, by the Central Funds Guiding the Local Science and Technology Development of Gansu Province, Grant No.~22ZY1QA006, by Gansu International Collaboration and Talents Recruitment Base of Particle Physics (2023-2027), by International Partnership Program of the Chinese Academy of Sciences, Grant No.~016GJHZ2022103FN, by National Natural Science Foundation of China, Grant No.~12375143, by National Key R\&D Program of China, Grant No.~2023YFA1606903 and by the Strategic Priority Research Program of the Chinese Academy of Sciences, Grant No.~XDB34000000.

\end{document}